# THE RELATION BEFORE LORENZ

Kirchhoff, 1857, and the symbols Lorenz inherited

María Florentina Mendoza Urán

ORCID 0009-0007-7882-6418

## Abstract

A relation between the quantities now called the scalar potential and the vector potential, of the same form, up to a sign, as the one the literature attributes to Ludvig Lorenz in 1867, was written ten years earlier by Gustav Kirchhoff, with the same status: derived, unnumbered and unnamed. This paper documents, on the basis of the 1857 originals, that Kirchhoff obtains the relation as the result of a calculation and does not impose it; that the quantity now called the vector potential bears no name at all in his text, whereas the scalar one bears the name "Potentialfunction", potential function, in both papers; that Lorenz writes those same letters, Ω, U, V and W, and inherits with them the absence of a name for the three components, although the relation he writes is in other quantities, α, β and γ, and carries the opposite sign; and that the English translation of one of the two papers suppresses the word "function" from the term "Potentialfunction", with which Kirchhoff names his scalar quantity. The paper concludes that the pattern documented in the 1867 paper predates Lorenz and comes from the source from which he starts: a relation without number or name, components without a denomination, and a vocabulary added later.

In the 1857 issue in which Kirchhoff publishes his paper on the motion of electricity in wires, the editor of the Annalen adds, at the end of the number, a remark signed by himself. He relates that Wilhelm Weber, during his recent stay in Berlin, showed him a paper of his own on the same subject, already fully worked out, which he did not yet intend to send to press because he wished to await the results of an experimental investigation undertaken jointly with Rudolf Kohlrausch. And he adds that Kirchhoff's passing through Berlin a few days later gave him the opportunity to talk with him about this coincidence, which he calls a pleasing one, because the two papers, starting from an essentially identical foundation, had reached concordant results in a subject as little explored as the laws of the formation of the current then were.[1]

Two men, separately, with the same result in hand, and an editor who puts it in writing. That result is what this article is about: what exactly is on the pages Kirchhoff published that year, in what words it is said, and what has been added to it since that is not on them.

For what is cited today from those pages is not exactly what is on them. A word is missing that a translator suppressed, the name of what was written there ended up being someone else's, and there is a surplus vocabulary that its author did not use. What follows is a path through those pages, in their own order.

Each of those three things has its place in what follows. The missing word, in section 6. The name that ended up being someone else's, in sections 8 and 10. The surplus vocabulary, in section 12.

## 1. Introduction

A previous investigation documented that the word potential does not appear in Lorenz's 1867 paper, that the magnetic field does not figure in it as a quantity of the theory, and that the vocabulary with which that paper is described today enters the literature in 1884, with Hertz.[2]

It remains to be established where that absence comes from. Lorenz does not build his equations from scratch: he starts from Kirchhoff's, and says so explicitly on the first page of his development.[3] The question, therefore, is whether the feature documented in 1867 is his own or whether he inherits it.

This paper examines the two articles Kirchhoff published in 1857 on the motion of electricity, in their German originals, and establishes what is in them and what is not.

## 2. The two papers of 1857

Kirchhoff published two articles that year with almost identical titles, and it is worth telling them apart, because they do not say the same thing.

The first, “Ueber die Bewegung der Elektricität in Drähten”, on the motion of electricity in wires, appeared in issue 2 of the Annalen der Physik und Chemie, volume 100 in Poggendorff's numbering, pages 193 to 217.[4] It deals with an infinitely thin wire and works in one dimension.

The second, “Ueber die Bewegung der Elektricität in Leitern”, on the motion of electricity in conductors, appeared in volume 102, pages 529 to 544.[5] It deals with conducting bodies in three dimensions.

The first was translated into English in the Philosophical Magazine, series 4, volume 13, pages 393 to 412, with the note “Translated from Poggendorff's Annalen, 1857, No. 2”.[6] The translation is unsigned. The second had no English translation until 1994, when Peter Graneau and André Koch Torres Assis published one in the journal Apeiron, presenting it as the first and basing it on the 1882 reprint.[7]

The one Lorenz cites, and from which he starts, is the second. On page 246 of his paper he writes: “Kirchhoff hat (Pogg. Ann. Bd. 102) die Gesetze für die Bewegung der Elektricität in den Körpern mit constantem Leitungsvermögen durch folgende Gleichungen ausgedrückt”, that is, that Kirchhoff has expressed by the following equations the laws of the motion of electricity in bodies of constant conductivity; and he then reproduces equations (1).[8]

## 3. Poggendorff's comment

In volume 100 of the Annalen, the comment Poggendorff adds to Kirchhoff's paper takes up two pages. The first bears the printed number 351. The second bears the printed number 252, and should bear 352.[9] The literature cites that text as “100, 351-352”, which is the correct location; the error is the printer's.

It is a minor detail, and it is recorded here for what it teaches: that pages have to be read in the volume, not in the citation. That is the criterion for everything that follows.

## 4. What Kirchhoff writes in the paper on wires

The article opens by stating its object. On the first line of page 193 Kirchhoff writes: “Ich habe versucht eine allgemeine Theorie der Bewegung der Elektricität in einem unendlich dünnen Drahte aufzustellen”, I have attempted to set up a general theory of the motion of electricity in an infinitely thin wire.

In it, Kirchhoff introduces two quantities.

The first is a scalar, and he names it. On page 194 of the German original he writes: “Es bezeichne V die Potentialfunction der freien Elektricität in Bezug auf den betrachteten Punkt, also die Summe der einzelnen freien Elektricitätsmengen, eine jede dividirt durch ihre Entfernung von dem Punkte”.[10] That is: let V denote the potential function of the free electricity with respect to the point considered, that is, the sum of the individual quantities of free electricity, each divided by its distance from the point.

The second he does not name. On page 199 he writes: “Es wird daher die ganze inducirte elektromotorische Kraft = $-(8/c^2)\ \partial W/\partial t$, wenn man setzt”, the whole induced electromotive force turns out to be minus eight over c squared times the time derivative of W, if one sets; and he then defines W as an integral of the current intensity over the wire.[11] He gives it no name; the count in section 11 confirms this.

He had introduced that induced electromotive force earlier, on page 197, “nach Weber”, according to Weber, with a footnote to the “Elektrodynamische Maassbestimmungen” of 1846, page 354, and of 1856, page 268, the latter by Weber and Kohlrausch. And in the same place he defines the constant:

"…und c die constante Geschwindigkeit, mit der zwei Elektricitätstheile gegen einander bewegt werden müssen, damit sie keine Kraft auf einander ausüben", and c the constant velocity with which two portions of electricity must move relative to each other in order to exert no force on each other. It is the same definition, in the same terms, that he will give on page 530 of the paper on conductors.

Before writing that equation he states an assumption. On the same page 199: "Bei einem stationären elektrischen Strome ist die Stromdichtigkeit gleich dem Producte aus der, auf die Einheit der Elektricitätsmenge bezogenen, elektromotorischen Kraft in die Leitungsfähigkeit; ich mache die Annahme, dass dasselbe auch stattfindet, wenn der Strom kein stationärer ist". In a steady current the current density is equal to the product of the electromotive force, referred to the unit quantity of electricity, and the conductivity; I make the assumption that the same holds when the current is not steady. And he gives the condition under which the assumption is fulfilled: that the time during which a particle of electricity continues to move by inertia, once the accelerating forces have ceased, may be regarded as infinitely small.

And on the same page 199 he writes the equation that relates both to the current density:

$$J = -2k\,(\partial V/\partial s + (4/c^2)\,\partial W/\partial t)$$

From that density he derives the intensity, multiplying by $\rho\, d\rho\, d\psi$ and integrating over the radius "von 0 bis α", from zero to α. And w, the quantity that appears in equation (2), is the one he defines on page 200 and develops in equation (3) on page 201: an integral extended over the whole wire. Equation (2) thus reads:

$$i = -2\pi k\alpha^2\,(\partial V/\partial s + (4/c^2)\,\partial w/\partial t)$$

On page 201 he adds a fourth equation: "Zu den Gleichungen 1, 2 und 3 zwischen den 4 Grössen i, e, V, w lässt sich noch eine vierte hinzufügen". To equations 1, 2 and 3 between the four quantities, a fourth can still be added. He obtains it by counting what enters and leaves an element of the wire: through one cross-section the quantity idt of positive electricity enters in the time dt, through the other $(i + \partial i/\partial s\, ds)dt$ leaves, the negative electricity moves in equal quantity in the opposite direction, and the free electricity of the element decreases by $2(\partial i/\partial s)ds\, dt$. Hence equation (4):

$$2\,\partial i/\partial s = -\partial e/\partial t$$

At the foot of that page he states what it rests on, and qualifies it in the same breath: "Der Ableitung dieser Gleichung liegt die Vorstellung zu Grunde, dass auch bei dem nicht stationären Strome durch jeden Querschnitt des Leiters gleichzeitig gleiche Mengen der beiden Elektricitäten in entgegengesetzter Richtung sich bewegen. Wollte man diese Vorstellung nicht festhalten, so würde aber die Gleichung doch noch gelten; man müsste dann nur die Stromintensität definiren als das arithmetische Mittel aus den Mengen beider Elektricitäten, welche in der Zeiteinheit durch den Querschnitt des Leiters in entgegengesetzter Richtung gehen". The derivation of this equation rests on the conception that, in a non-steady current too, equal quantities of the two electricities move simultaneously through each cross-section in opposite directions. Should one not wish to hold on to this conception, the equation would nevertheless still hold; one would then only have to define the current intensity as the arithmetic mean of the quantities of the two electricities that pass in unit time through the cross-section in opposite directions. The footnote continues at the foot of page 202.

Comparison with the 1882 reprint of his collected works shows no differences in these passages.[12]

## 5. The velocity of propagation in the paper on wires

There is a result in the first paper that is worth recording, because it is obtained without recourse to any property of the medium outside the wire.

On pages 206 to 208 of the original, Kirchhoff solves the equations he has set up and obtains a wave solution: for a wire that returns on itself, the periodic solutions carry the factor $m(2\pi/l)(c/\sqrt{2})t$, with m an integer, and the velocity of propagation turns out to be c divided by the square root of two.[13]

And he himself states the result and comments on it. On pages 209 and 210 of the German original he writes: “Die Fortpflanzungsgeschwindigkeit einer elektrischen Welle hat sich hier = $c/\sqrt{2}$ ergeben, also als unabhängig sowohl von dem Querschnitt, als von der Leitungsfähigkeit des Drahtes, als endlich von der Dichtigkeit der Elektricität; ihr Werth ist der von 41950 Meilen in einer Secunde, also sehr nahe gleich der Geschwindigkeit des Lichtes im leeren Raume”. That is: the velocity of propagation of an electric wave has turned out here to be equal to c divided by the square root of two, that is, independent both of the cross-section and of the conductivity of the wire, and finally of the density of the electricity; its value is that of 41,950 miles in one second, that is, very nearly equal to the velocity of light in empty space.[14]

The mile is the German mile of the time. Kirchhoff's own figures say so: his c, $4.39 \times 10^{11}$ millimetres per second on page 205, divided by the square root of two, gives some 310,000 kilometres per second, and that is 41,950 miles of about 7.4 kilometres each.

The c that appears there is the constant of Weber's law of induction, the same one that enters the equations of sections 4 and 7. For its value, Kirchhoff refers to the “Elektrodynamische Maassbestimmungen” of 1846, by Weber, and of 1856, by Weber and Kohlrausch, and to Weber's measurement of Jacobi's standard wire.[15]

So the velocity with which the disturbance travels along the wire is obtained from a constant measured between quantities of electricity, and not from any quantity that characterizes the medium surrounding the conductor.

It is worth seeing how he gets there. The path can be followed in the German original. On page 203 he closes the system: after substituting V and w, and calling $\gamma$ the logarithm and r the resistance of the whole wire, he obtains $i = -4\gamma\,(l/r)(\partial e/\partial s + (4/c^2)\,\partial i/\partial t)$, and writes: “Aus dieser Gleichung, in Verbindung mit der Gleichung 4, nämlich $2\,\partial i/\partial s = -\partial e/\partial t$, sind i und e als Functionen von s und t zu bestimmen”. From this equation, in conjunction with equation 4, i and e are to be determined as functions of s and t.

On page 204 he tries $e = X \sin ns$, $i = Y \cos ns$, eliminates Y and obtains:

$$d^2X/dt^2 + (c^2r/16\gamma l)\, dX/dt + (c^2n^2/2)\, X = 0$$

And on that same page, to decide whether the roots are real or imaginary, he turns to a measured conductor: Jacobi's resistance standard, whose resistance Weber had measured. On page 205 he gives $c = 4.39 \cdot 10^{11}$ and $r = 2.482 \cdot 10^{-13}$, and from them $32\gamma/(rc\sqrt{2}) = 2070$; since that quantity is very large, the roots turn out to be imaginary and the solutions periodic.

On page 209 he writes the passage that gives his reading of the result: “Der Ausdruck von e zeigt eine sehr merkwürdige Analogie zwischen der Fortpflanzung der Elektricität in dem Drahte und der Fortpflanzung einer Welle in einer gespannten Saite oder einem longitudinal schwingenden elastischen Stabe”. The expression for e shows a very remarkable analogy between the propagation of electricity in the wire and the propagation of a wave in a stretched string or in an elastic rod vibrating longitudinally. And in stating the decomposition he introduces a caution about his own vocabulary: “so zerfällt, wie ich mich ausdrücken will, die Elektricität in zwei Wellen von gleicher Stärke”, the electricity splits, if I may put it that way, into two waves of equal strength.

On page 210 he returns to that analogy to settle the conditions at the ends, again with caution: “Es entspricht also in gewisser Hinsicht das abgeleitete Ende dem festen Ende eines longitudinal schwingenden Stabes, das isolirte dem freien Ende des Stabes”. The earthed end corresponds, in a certain respect, to the fixed end of a longitudinally vibrating rod, and the insulated end to the free end.

It is worth recording what is not there. Nowhere in this paper does Kirchhoff write a wave equation as a single equation: he works with two first-order equations, the one on page 203 and equation (4). Nor does he use the expression electromagnetic wave. What he writes is a second-order differential equation with a damping term, a solution that has the form of a travelling wave, and a declared analogy with two mechanical systems.

Ten years later, Lorenz will put the identity between the vibrations of light and electric currents into the title of his paper. Kirchhoff, in 1857, has the number in front of him, compares it with that of light, leaves it at "sehr nahe gleich", very nearly equal, and then goes on to speak of a stretched string and an elastic rod.

## 6. A suppression in the English translation

Before the word, a number. On page 402 of the English translation, the length of Jacobi's standard wire is given as "7·620 inches". The German original, on page 204, reads "$7^{m}$,620": metres, with the unit as a superscript and the comma as the decimal separator. The radius, "$0^{mm}$,333", comes through correctly. Kirchhoff's own calculation decides between the two readings: γ, the logarithm of the length divided by the radius, is according to him "sehr nahe = 10", and with 7620 millimetres it is 10.04, whereas with 7.620 inches it would be 6.37. In the paper on conductors, on page 543, he again gives the measurement in metres, "$7^{m}$,62". So the translator changed a unit on page 402. And on page 393, the first page of the translation, he changed a word.

The English version of that first paper renders the passage from page 194 as follows: "Let V be the potential of the free electricity with reference to the point under consideration; that is to say, the sum of all the single quantities of free electricity, each divided by its distance from the point".[16]

The definition is preserved in full. What disappears is a word: where Kirchhoff writes "Potentialfunction", potential function, the translation has "potential".

The suppression is not forced by the language. The term "potential function" existed in English: it is the one George Green coins in 1828, on the first page of his essay, when he writes that henceforth, for abridgment, he will call that function "the potential function arising from the system S". The translator had it at his disposal.[17]

The other English translation of Kirchhoff, the one Peter Graneau and André Koch Torres Assis published in 1994 for the paper on conductors, does keep the word: where Kirchhoff writes "Potentialfunction", they write "potential function of the free electricity", and they leave U, V and W without a denomination, as in the original.[18]

What changes is not what the equation calculates, but the status of what it names. "Potential function" says that it is a function, an object constructed in order to calculate; the noun is "function", and "potential" qualifies it. "Potential" turns the qualifier into a noun: it is no longer a function of a certain kind, but a thing that one has. And of a thing one may ask whether it is physical.

## 7. What Kirchhoff writes in the paper on conductors

The second article, the one Lorenz cites, opens by announcing its purpose: "ich will jetzt zeigen, wie die dort durchgeführten Betrachtungen sich so verallgemeinern lassen, dass sie auf Leiter jeder Gestalt anwendbar werden", I now wish to show how the considerations developed there can be generalized so as to become applicable to conductors of any shape. In it there are four quantities. The scalar one bears a name; the three components do not.

On page 530 Kirchhoff writes the three volume integrals U, V and W, and then: "wo die Integrationen über das ganze Volumen des Leiters ausgedehnt gedacht sind. Nach dem Weber'schen Gesetze der Induction sind dann die Componenten des zweiten Theiles der betrachteten elektromotorischen Kraft", where the integrations are understood to extend over the whole volume of the conductor; according to Weber's law of induction, the components of the second part of the electromotive force under consideration are then.[19] He does not name them.

On that same page he defines the constant that appears in those components: "wo c diejenige constante Geschwindigkeit bezeichnet, mit der zwei Elektricitätstheile gegen einander bewegt werden müssen, damit sie keine Kraft auf einander ausüben", where c denotes that constant velocity with which two portions of electricity must move relative to each other in order to exert no force on each other. That is the definition of c, that of the constant of Weber's law of induction, and not that of the velocity of light. Its value is another matter: in the paper on wires, Kirchhoff points out that c divided by the square root of two turns out to be very nearly the velocity of light in empty space, as recorded in section 5.

And on that same page he writes the three equations:

$$u = -2k\,(\partial\Omega/\partial x + (4/c^2)\,\partial U/\partial t)$$

$$v = -2k\,(\partial\Omega/\partial y + (4/c^2)\,\partial V/\partial t)$$

$$w = -2k\,(\partial\Omega/\partial z + (4/c^2)\,\partial W/\partial t)$$

The scalar quantity Ω appears defined in equation (4) on the same page, introduced with "dann ist", then is.[20] And he does name it, though not on that page but on the previous one. On page 529, listing the two parts of the electromotive force, he writes: "Bezeichnet Ω die Potentialfunction der freien Elektricität in Bezug auf den Punkt (x, y, z), so sind die Componenten des ersten Theiles der betrachteten elektromotorischen Kraft $-2\ \partial\Omega/\partial x$, $-2\ \partial\Omega/\partial y$, $-2\ \partial\Omega/\partial z$". Let Ω denote the potential function of the free electricity with respect to the point (x, y, z). It is the same compound he uses in the paper on wires.

## 8. The relation, in 1857

Here, up to a sign, is the relation the literature attributes to Lorenz ten years later. It is worth seeing at what point it appears, and how.

On page 531 he first writes the two equations that express the changes over time of the free electricity. For each point of the interior, equation (5):

$$\partial u/\partial x + \partial v/\partial y + \partial w/\partial z = -\tfrac{1}{2}\,\partial \varepsilon/\partial t$$

And for each point of the surface, calling N the normal directed towards the interior, equation (6):

$$u \cos(N, x) + v \cos(N, y) + w \cos(N, z) = -\tfrac{1}{2}\,\partial e/\partial t$$

Neither of them contains any physical constant, neither c nor k, just as equation (4) of the paper on wires contains none.

And then, on that same page, he announces: "Aus den aufgestellten Gleichungen lässt sich eine merkwürdige Relation zwischen ε und Ω herleiten", from the equations set up, a remarkable relation between ε and Ω can be derived.[21]

The development continues on page 532. There he writes that one convinces oneself of the smallness of certain terms "durch dasselbe Verfahren, durch welches Gauss nachgewiesen hat", by the same procedure by which Gauss proved that, in the potential of masses filling a space continuously, the masses infinitely close to the point contribute infinitely little; and he obtains the final expression by integrating by parts.[22] And on the first line of page 533 he writes the relation:

$$\partial U/\partial x + \partial V/\partial y + \partial W/\partial z = \tfrac{1}{2}\,\partial\Omega/\partial t$$

Next, with "Hieraus folgt", hence it follows, he obtains equation (7): $\partial\varepsilon/\partial t = -8k\,(2\pi\varepsilon - (1/c^2)\,\partial^2\Omega/\partial t^2)$. That is the remarkable relation between ε and Ω that he had announced on page 531. The relation between U, V, W and Ω is the intermediate step towards it, and it is unnumbered.[23]

It is worth making clear in what terms it appears. Kirchhoff obtains it: he derives it from the equations he already has, writes it unnumbered as an intermediate step, and goes on calculating. He does not state it as a condition on quantities previously declared to be undetermined, he does not mention that there are other quantities producing the same results, and he does not choose between alternatives. It is a step in a derivation.

It is, therefore, the same case as the one documented for the 1867 paper: a derived relation, not an imposed condition. And it is written ten years earlier.

That line, the first on page 533, unnumbered, is the one that will receive a name one hundred and forty-nine years later. Section 12 will show who gives it that name, and when.

## 9. The short wire and the long wire

In the second part of the paper on conductors, Kirchhoff applies his general theory to the case he had treated in the first paper, that of a cylindrical wire of infinitely small cross-section. On page 538 he writes that the equations he arrives at are the same as those of the earlier paper: “Die Gleichungen (14) und (15) sind aber dieselben als diejenigen, zu denen ich für denselben Fall in der oben angeführten Abhandlung gekommen bin”.

On page 539 he summarizes what he had shown there: “Ich habe die Lösung der Gleichungen (14) und (15) an dem mehrfach erwähnten Orte für einen Fall discutirt, dem man sich um so mehr nähert, je kleiner man den Widerstand des Drahtes macht, und habe nachgewiesen, dass in diesem Falle sich die Elektricität in dem Drahte ähnlich fortpflanzt, wie eine Welle in einer gespannten Saite, und zwar mit der Geschwindigkeit, die das Licht im leeren Raume hat”. I have discussed the solution of equations (14) and (15) in the place mentioned several times, for a case that is approached the more closely the smaller the resistance of the wire is made, and have shown that in that case the electricity propagates in the wire in a manner similar to a wave in a stretched string, and indeed with the velocity that light has in empty space.

It is the same result as on pages 209 and 210 of the paper on wires, put another way. There the value was left at “sehr nahe gleich”, very nearly equal, to the velocity of light in empty space. Here it is “die Geschwindigkeit, die das Licht im leeren Raume hat”, the velocity that light has.

He then treats the opposite case, the one that arises the greater the resistance of the wire. On page 541 he writes that in the earlier paper he had regarded the quantity $32\gamma/(cr\sqrt{2})$ as infinitely large, and that he now supposes it to be infinitely small; the two roots $\lambda_1$ and $\lambda_2$ are then real.

On page 542, after writing solutions (16) and (17), he says: “Diese Ausdrücke sind unabhängig von c”, these expressions are independent of c. And on page 543 he completes the sentence: they are the solutions of the equations obtained from (14) and (15) “wenn man c unendlich gross setzt”, when c is made infinitely large. Eliminating i, he is left with:

$$\partial E/\partial t = (8\gamma l/r)\, \partial^2 E/\partial x^2$$

And he describes it himself: “welche Gleichung von derselben Form als diejenige ist, welche die Fortpflanzung der geleiteten Wärme bestimmt. In dem betrachteten Falle pflanzt sich also die Elektricität ähnlich wie die geleitete Wärme fort”. An equation of the same form as the one that determines the propagation of conducted heat; in the case considered, the electricity therefore propagates in a manner similar to conducted heat.

On page 543 he compares the two cases with a real wire. For Jacobi's standard the quantity $32\gamma/(rc\sqrt{2})$ is 2070; for a wire of the same material and the same cross-section, 1000 kilometres long, it is 0.034. And on page 544: “sie kann bei jenem näherungsweise als unendlich gross, bei diesem als unendlich klein betrachtet werden; es pflanzt sich bei jenem die Elektricität ähnlich wie eine Welle in einer Saite, bei diesem wie die geleitete Wärme fort”. In the former it can be regarded approximately as infinitely large, in the latter as infinitely small; in the former the electricity propagates in a manner similar to a wave in a string, in the latter like conducted heat.

The material is the same and the cross-section is the same. What changes is the length, and with it the law that the electricity follows.

The paper ends with Thomson. Kirchhoff writes that Thomson studied the motion of electricity in a submarine telegraph cable, that to do so he made the assumption, “ohne die Zulässigkeit derselben zu prüfen”, without checking its admissibility, that induction exerts no appreciable effect, and that he showed that the electricity then moves like conducted heat. And he adds that the considerations of his own paper show that this assumption is already fulfilled in a simple wire if its length is great enough, and that it will be all the more correct in a submarine telegraph cable, in which the motion of the electricity is considerably retarded by conduction through the sea water.[24]

The footnote with which Kirchhoff refers to Thomson gives “Phil. Mag. Ser. IV, Vol. II, p. 157”. That page of that volume does not contain Thomson's paper, but a Royal Society abstract on another subject. The paper is the one in the Proceedings of the Royal Society of 1855, received on 3 May, and it is by that original that it is given in the references.

In the passages of this paper where he makes the comparison with a wave, the one on page 539 and the one on page 544, the wording is the same: on page 539 he writes “ähnlich fortpflanzt, wie”, propagates in a manner similar to, and on page 544, “ähnlich wie”.

## 10. What Lorenz inherits

The comparison of the symbols is direct. Kirchhoff writes Ω for the scalar quantity and U, V, W for the three components. Lorenz writes Ω for the scalar quantity and U, V, W for the three components.[25] They are the same letters.

What Lorenz adds is retardation. His $\bar{\Omega}$ is Kirchhoff's Ω with the density evaluated at the instant $t - r/a$. His α, β and γ, on the other hand, are not Kirchhoff's U, V and W retarded: he himself writes, on page 249, that his equations differ from (1) in that, instead of U, V, W, they contain the somewhat less compound terms α, β and γ. And on page 250 he specifies that, if a were taken to be infinitely large, the equation would take the form that results from Neumann's electrodynamic theory. And he declares as much: he introduces $\bar{\Omega}$ as “eine neue Function”, a new function, and α, β, γ “der Kürze wegen”, for brevity.[26]

Nor is the relation Lorenz writes identical. He writes it for the first time on page 253, when he obtains by partial integration, with the designations α, β and γ, the derivative of $\bar{\Omega}$ as equal to minus twice the sum of the derivatives of α, β and γ; and he recalls it on page 259, “in Uebereinstimmung mit der frühern Bezeichnung”, in accordance with the earlier notation, now as the sum of those derivatives equal to minus one half of the time derivative of $\bar{\Omega}$, whereas Kirchhoff, on his page 533, had written it as equal to plus one half of the derivative of Ω. And on page 256 he expressly departs from him: he writes that his result differs from the one Kirchhoff deduced from equations (1), namely, that in general there is free electricity in the interior of a conductor.

So the absence of a name for the three components in the 1867 paper is not a peculiarity of Lorenz: it is what was in his source. With the scalar quantity the opposite happens: Kirchhoff calls it “Potentialfunction”, and Lorenz does not carry that word over.

A relation of the same form as the one the literature attributes to 1867 is written in 1857, with Kirchhoff's letters and with Kirchhoff's sign.

## 11. The symbols and what they name

A remark on the vocabulary of the two papers, worth making because it affects the reading of both.

In the paper on wires, Kirchhoff uses J for the current density and i for the intensity. These are the quantities that are measured. The quantity that enters into induction is W, and it is an integral of the current intensity extended over the wire. In the paper on conductors, the measured quantities are u, v and w, the components of the current density, and those that enter into induction are U, V and W, integrals of those same components over the volume of the conductor.[27]

So, in both papers, the quantity now called the vector potential is, in the text, an integral of current. An integral of the current, weighted by the distances, which Kirchhoff introduces in order to write the components of the electromotive force with it. And its author gives it no other name.

A count over the paper on wires confirms this from the opposite side. In its twenty-five pages, the root “magnet” appears only once, on page 204, in the expression “nach elektromagnetischem Maasse”, in electromagnetic measure, referring to the resistance of the standard wire that Weber measured.[28] The word “Potential” appears three times, and all three refer to V, the scalar quantity: in its definition,

“Potentialfunction”, on page 194, and on pages 210 and 211, where it vanishes owing to the conditions at the ends of the wire.[29] None of them refers either to W or to w. And neither Neumann nor Maxwell appears: as a theoretical source only Weber appears, five times, and alongside him Jacobi and Kohlrausch, in the notes on measurement.

## 12. What the literature says

It is worth seeing now how these pages are described today.

Jackson and Okun write that Kirchhoff was the first to write the vector potential explicitly, in component form, and they refer to page 530 of the second paper. They add that he wrote the components of the induced current density as the conductivity multiplied by the negative of the sum of the gradient of the scalar potential and the time derivative of the vector potential. And they place on pages 532 and 533 the first published relation between potentials in what we now know as a particular gauge.[30]

The location is exact: those are the pages, and the equations are there. What is added is the vocabulary. On page 530 there is no vector potential: there are three unnamed integrals. On pages 532 and 533 there is no gauge: there is a derived, unnumbered relation, which is a step towards equation (7).

This is not an error. It is the rendering into present-day language of a text that does not use that language, and as such it serves its purpose. What is worth pointing out is that the text does not use it, and that the reader who comes to the text through that description finds in it a terminology its author did not put there.

The entry of that vocabulary can be traced, and the authors themselves declare it. Jackson and Okun note in their introduction that “the word ‘gauge’ was not used in English for transformations such as (1,a,b,c) until 1929 (Weyl, 1929a). It is convenient, nevertheless, to use the modern terminology even when discussing the works of 19th century physicists”. It is Jackson and Okun, in 2001, who describe the relation on pages 532 and 533 as the first published relation between potentials in what is now known as a particular gauge, without giving it a name. The name is proposed by José Antonio Heras in 2006: in “The Kirchhoff gauge” he writes, of the condition he takes from them, “which will be called here the Kirchhoff condition”, and in a later paper he confirms it: “the present author has proposed to call Eq. (6) the Kirchhoff gauge”. It is the name McDonald records in his note, referring to the 2006 paper by Heras.[31]

What follows are three readings of the result of the paper on wires, set out in order of publication, each in its author's own words.

All three start from the same pages, and they do not say the same thing about them. They are set down without being discussed, because what matters here is the distance between them, and between all of them and the text.

Kirk T. McDonald writes that in 1857 Kirchhoff published two papers on the motion of electricity in wires, based on Weber's action-at-a-distance electrodynamics; that considering a single wire followed the telegraphic practice of the time, in which only a single wire appeared to be involved and the role of the earth as a return path closing the circuit was not yet recognized, nor the capacitance between the wire and the earth, nor the self-inductance of the circuit formed by the two; and that the derivation of the “telegrapher's equation” for a two-wire line found in textbooks today, by means of Kirchhoff's circuit laws, was first given by Heaviside in 1876.[32]

In that same note he points out that, since Weber's electrodynamics was one of action at a distance and not a field theory in Faraday's sense, neither Weber nor Kirchhoff inferred that light must be an electromagnetic phenomenon. He holds that Kirchhoff's argument was the first demonstration that the velocity of surface waves of charge and current coincides with the velocity of waves in the medium outside the conductor. And he adds that one ingredient of his analysis was the condition between potentials now called the Kirchhoff gauge, the first use of a gauge condition.[33]

André Koch Torres Assis writes that Kirchhoff and Weber were the first to deduce the telegraphy equation taking into account the self-inductance of the wire, that both worked with Weber's electrodynamics, and that they showed, in particular, that when the resistance of the wire was negligible the electric wave propagates with the velocity of light. In a footnote he puts it more cautiously: that the velocity of propagation of the electric wave is very nearly equal to the velocity of light in vacuum.[34]

He adds that in 1833 Gauss and Weber invented the world's first operational electromagnetic telegraph, a three-kilometre double cable linking the University of Göttingen, where Weber was professor of physics, with the astronomical observatory directed by Gauss, and that it worked according to Faraday's law of induction, discovered two years earlier. And that twenty-four years later, in 1857, Weber and Kirchhoff were the first to derive the complete telegraphy equation theoretically; that, as Poggendorff points out in his comment, they worked independently of each other and arrived at the same result at the same time; and that, in the concepts and terminology usual today in circuit theory, they were the first to take into account not only the capacitance and the resistance of the wire, but also its self-inductance. He specifies that they derived it using Weber's force law of 1846. And he warns, in another note, that Weber's constant c, which Kohlrausch measured with him, must not be confused with the constant c of today's textbooks.[35]

Frederick David Tombe writes that Kirchhoff is credited with the telegrapher's equations, mathematically equivalent to Maxwell's electromagnetic wave equations, and that they differ only in the physical context in which they apply: Kirchhoff believed they held for the propagation of electric signals along a conducting wire, and Maxwell for wireless radiation in space, which he considered densely packed with tiny aethereal vortices. And he holds that Kirchhoff's error seems to have been to merge capacitance and inductance on the laboratory scale when deriving the wave equations, without distinguishing the electric field terms induced by capacitance from those arising from time-varying electromagnetic induction.[36]

In an earlier paper he is more precise. He writes that Kirchhoff was only a few steps away from obtaining the telegrapher's equations that Oliver Heaviside would obtain about thirty years later, and that he was working with Weber's equations, where Weber's constant, and with it the speed of light, were firmly planted in the magnetic domain; and that his error was to criss-cross, in his analysis, the electrostatic field induced by the capacitance with the electric field induced by self-inductance, as though they were equivalent for substitution purposes. And he adds that Heaviside did something similar: that in purporting to derive the telegrapher's equations he in fact, without realizing it, derived the wireless electromagnetic wave equations, with the same substitution that textbooks use today. He also records that Weber and Kirchhoff subscribed to the Fechner hypothesis.[37] He further specifies that Kirchhoff's context was that of the propagation of a time-varying electric current in a wire, but that he wrongly linked the propagation speed to the two reactive transients, capacitance and self-inductance.[38]

None of the three readings alters what is on page 533: a derived relation, written unnumbered, between a named potential function and three unnamed integrals of current.

## 13. Conclusions

One. The relation between the quantities now called the scalar potential and the vector potential is written by Kirchhoff in 1857, on page 533 of volume 102 of the Annalen, ten years before Lorenz, and with the sign Kirchhoff gives it.

Two. Kirchhoff obtains it as the result of a derivation and writes it unnumbered, as an intermediate step towards equation (7), which is the one he announces as a remarkable relation. He does not impose it, he declares no prior indeterminacy, and he does not choose between alternatives.

Three. In the paper from which Lorenz starts, the three components bear no name: they are introduced with “wo die Integrationen … ausgedehnt gedacht sind”, where the integrations are understood to extend over the volume of the conductor. The scalar quantity does bear one: Kirchhoff calls it “Potentialfunction” on page 529, and reintroduces it on page 530 with “then is”.

Four. Lorenz reproduces Kirchhoff's letters and his equations; but the relation he writes is in other quantities, α, β and γ, which in his own words are less compound, and it carries the opposite sign. What he inherits without a name are the three components; with the scalar quantity the reverse is true, because the word was in his source and he does not carry it over.

Five. The scalar quantity bears the same name in both papers, “Potentialfunction”, potential function. And the English translation of the paper on wires suppresses the word “function”.

Six. In both papers, the quantity now called the vector potential is, in the text, an integral of current.

Seven. The vocabulary of gauge could not be in these texts: the word is not used in English for those transformations until 1929, as Jackson and Okun point out, while stating that they use it for convenience in describing nineteenth-century physicists. The first description of these pages in those terms appears to be that of Jackson and Okun, in 2001, and the name Kirchhoff gauge is proposed by Heras in 2006. It remains to be established, and is not claimed here, at what moment and by what route the term vector potential was brought into them.

Poggendorff put in writing, in 1857, that two men had arrived separately at the same result. What this paper puts in writing is what that result says on the pages where it was published, and what has been laid on top of it since: one word fewer, since the English translation of the same year; the description in terms of gauge, which appears to date from 2001; and the name Kirchhoff gauge, since 2006. And what remains to be known: when, and by what route, the term vector potential entered those pages.

## Notes

1. J. C. Poggendorff, “Bemerkung zu dem Aufsatz des Hrn. Prof. Kirchhoff”, Annalen der Physik und Chemie 100, 351-352 (1857), item XVIII of the issue. The second page bears the printed number 252 owing to a printer's error.

2. M. F. Mendoza Urán, “Lorenz vs Lorentz. La controversia estudiada a partir de los textos originales”, DOI 10.17605/OSF.IO/2WYG8 (2026), sections 3, 5 and 8.

3. L. Lorenz, “Ueber die Identität der Schwingungen des Lichts mit den elektrischen Strömen”, Annalen der Physik und Chemie 131, 243-263 (1867), p. 246.

4. G. Kirchhoff, “II. Ueber die Bewegung der Elektricität in Drähten”, Annalen der Physik und Chemie, Band 176 in the continuous numbering, Pogg. Ann. 100, no. 2, 193-217 (1857). Reprinted in Gesammelte Abhandlungen von G. Kirchhoff, J. A. Barth, Leipzig, 1882, 131-154. DOI 10.1002/andp.18571760203.

5. G. Kirchhoff, “II. Ueber die Bewegung der Elektricität in Leitern”, Annalen der Physik und Chemie, Band 178 in the continuous numbering, Pogg. Ann. 102, Heft 12, 529-544 (1857). DOI 10.1002/andp.18571781203. Reprinted in Gesammelte Abhandlungen, 154-168.

6. G. Kirchhoff, “LIV. On the Motion of Electricity in Wires”, The London, Edinburgh and Dublin Philosophical Magazine and Journal of Science, series 4, vol. 13, no. 88, June 1857, 393-412. The header note is on p. 393; the length of Jacobi's standard wire, on p. 402.

7. P. Graneau and A. K. T. Assis, “Kirchhoff on the Motion of Electricity in Conductors”, Apeiron 19, 19-25 (1994), p. 19.

8. Lorenz 1867, p. 246. The original is in Danish: “Om Identiteten af Lyssvingninger og elektriske Strømme”, Oversigt over det Kongelige Danske Videnskabernes Selskabs Forhandlinger, 1867, no. 1, pp. 26-45, communicated to the Academy on 25 January 1866. The header of the German version, on p. 243, states that it comes from that bulletin and that it was sent by the author himself, “vom Hrn. Verfasser übersandt”. The two versions differ in two places that bear on what is cited here: where the German says “the laws of the motion of electricity”, the Danish, on its p. 28, says “the laws of the electric currents”; and the “der Kürze wegen”, for brevity, on p. 249 of the German is not in the Danish, which introduces α, β and γ without that phrase. Neither text records the reason for the differences.

9. Annalen der Physik und Chemie, Pogg. Ann. 100 (1857). The second page of the comment bears the printed number “252”. Verified on the scan of the volume.

10. Kirchhoff 1857 (Drähten), German original, p. 194. Verified on the scan of the volume. The passage coincides with the 1882 reprint, p. 132.

11. Ibid., p. 199. It coincides with the 1882 reprint, pp. 136-137.